\documentclass[prl,article,twocolumn,preprintnumbers,superscriptaddress,amsmath,amssymb,aps,longbibliography]{revtex4-1}

\usepackage{graphicx}
\usepackage{dcolumn}
\usepackage{bm}
\usepackage{multirow} 
\usepackage{comment}
\usepackage{url}
\usepackage{epsfig}
\usepackage{amsfonts}
\usepackage{amsmath}
\usepackage{bm}
\usepackage{graphicx}
\usepackage{color}
\usepackage{amssymb,amsmath,hyperref,slashed,color,graphicx,bm,cancel,url}
\usepackage[normalem]{ulem}
\usepackage[T1]{fontenc}
\usepackage[dvipsnames,table]{xcolor}

\hypersetup{colorlinks,
			citecolor      = niceblue, 
			linkcolor       = niceblue, 
			urlcolor        = niceblue, 
			anchorcolor = niceblue}
\definecolor{niceblue}{rgb}{0.1,0.2,0.6}
\hypersetup{breaklinks=true}

\newcommand{\be}{\begin{equation}}
\newcommand{\ee}{\end{equation}}
\newcommand{\bea}{\begin{eqnarray}}
\newcommand{\eea}{\end{eqnarray}}

\newcommand{\gagg}{g_{a\gamma\gamma}}
\newcommand{\gae}{g_{ae}}
\newcommand{\agg}{a\gamma\gamma}

\begin{document}

\title{New solar X-ray constraints on keV Axion-Like Particles}

\author{Cyprien Beaufort}
\email{cyprien.beaufort@lpsc.in2p3.fr}
\affiliation{Laboratoire de Physique Subatomique et de Cosmologie, Universit\'{e} Grenoble-Alpes, CNRS/IN2P3, 38000 Grenoble, France}

\author{Mar Bastero-Gil}
\email{mbg@ugr.es}
\affiliation{Departamento de F{\'i}sica Te{\'o}rica y del Cosmos, Universidad de Granada, Granada-18071, Spain}
\affiliation{Laboratoire de Physique Subatomique et de Cosmologie, Universit\'{e} Grenoble-Alpes, CNRS/IN2P3, 38000 Grenoble, France}

\author{Tiffany Luce}
\email{tiffany.luce@physik.uni-freiburg.de}
\affiliation{Laboratoire de Physique Subatomique et de Cosmologie, Universit\'{e} Grenoble-Alpes, CNRS/IN2P3, 38000 Grenoble, France}

\author{Daniel Santos}
\email{daniel.santos@lpsc.in2p3.fr}
\affiliation{Laboratoire de Physique Subatomique et de Cosmologie, Universit\'{e} Grenoble-Alpes, CNRS/IN2P3, 38000 Grenoble, France}

\date{\today}

\begin{abstract}
The decay of Axion-Like Particles (ALPs) trapped in the solar gravitational field would contribute to the observed solar X-ray flux, hence constraining ALP models. We improve by one order of magnitude the existing limits in the parameter space $(g_{a\gamma\gamma}, m)$ by considering ALPs production via photon coalescence. For $g_{ae} \neq 0$, we demonstrate that trapped ALPs can be Compton-absorbed while crossing the Sun, resulting in two regimes in the exclusion limits, with a transition triggered by $g_{ae}$. Out of the transitional region, the solar X-ray constraints on ALPs are exclusively governed by $g_{a\gamma\gamma}$.

\end{abstract}

\maketitle

\emph{Introduction.---} ALPs are hypothetical massive bosons predicted in several extensions of the Standard Model and frequently introduced to address current puzzles such as dark matter, baryon asymmetry, or naturalness problems \cite{Arias2012, Ringwald2014, Choi2020, Irastorza2021}.

While ALPs with masses in the range (keV-MeV) are significantly constrained by cosmological bounds \cite{Cadamuro2012, Ringwald2012, ThesisCyprien}, they have received increasing interest \cite{Thorpe-Morgan2020, Baumholzer2020, Jaeckel2016, Dolan2017} since such heavy ALPs could play a role in the abundance of light elements produced during the Big Bang Nucleosynthesis \cite{Depta2020}, could shine a light on a measured emission line \cite{Jaeckel2014, Higaki2014} or could explain a resonance observed in particular nuclear transitions in atoms \cite{Ellwanger2016, Krasznahorkay2019}.  

ALPs would be produced in the Sun up to a kinematic limit of a few tens of keV, set by the temperature in the solar interior. A fraction of them would be sufficiently non-relativistic (NR) to be trapped in the solar gravitational field, then orbiting the Sun and accumulating over cosmic times \cite{DiLella2003}. We call them \textit{trapped} ALPs. 

A trapped ALP of mass $m$ can decay into two photons of energy $E_\gamma = m/2$, thus contributing to the observed solar luminosity. The requirement that the ALP-induced photons flux should not exceed the solar X-ray measurements is then a powerful way of constraining ALP models, as pointed out more than ten years ago \cite{Zioutas2004, Hannah2007, Hannah2010}.

Recently, DeRocco~\textit{et al.} revised the constraints on trapped ALPs \cite{DeRocco2022} based on the theoretical framework of \textit{solar basins} \cite{VanTilburg2020}. This framework has been developed at the same time as our theoretical description of trapped axions \cite{KKaxions}. While our model concerns higher-dimensional axions, the standard ALP case can be recovered by setting the extra-dimensional parameters to $\delta =1$ and $2 R=1~\rm{keV^{-1}}$. 

In this letter, we revisit the solar X-ray constraints on ALPs. Besides the use of an independent framework, the two main extensions concern:
\begin{itemize}
    \item the introduction of an additional production mechanism, the {\it photon coalescence}, which dominates the production of trapped hadronic ALPs;
    \item the {\it Compton absorption} of trapped ALPs crossing the Sun in their orbits, which partially counter-balances the production.
\end{itemize}

In this work, we use the solar X-ray data of SphinX in the range $E_\gamma \in [1.5, 6~\rm{keV}]$ measured during a deep solar minimum \cite{Sylwester2012, Sylwester2019}, as well as the data exploited by DeRocco~\textit{et al.} in the range $E_\gamma \in [3, 20~\rm{keV}]$ collected by the NuSTAR telescope \cite{NuSTAR2013}.

\vspace{0.4cm}


\emph{Theoretical description of trapped ALPs.---} We restrain our study to the case of pseudoscalar ALPs, hereafter denoted \textit{axions}. At tree level, the axions necessarily interact with photons, parameterized by the coupling $\gagg$ and, for non-hadronic models, the axions can also interact with electrons through the axion-electron coupling $\gae$. 

The two main axion-photon production mechanisms are the Primakoff process and the photon coalescence. The photon coalescence rate is given by the axion decay width $\Gamma_{\agg}$:
\be
\Gamma^{\rm{Coal.}} ~ \simeq \Gamma_{\agg} = \frac{\gagg^2 m^3}{64\pi}~,\label{eq:ProdCoal} 
\ee
where $m$ is the axion mass, while Primakoff production would be suppressed at small axion velocities \cite{DiLella2000}, in particular for trapped NR axions. However, as noticed in \cite{DeRocco2022}, once thermal effects are taken into account, the next leading term for Primakoff production is given at $O(v_e^2)$, with $v_e \simeq \sqrt{3 T/m_e}$  being the electron velocity and $T$ the solar temperature. Therefore, we obtain the Primakoff production rate:
\be
\Gamma^{\rm{Prim.}} ~\simeq~ \gagg^2 ~\frac{3 \alpha n_e T}{2 m_e} \,,
\ee
where $\alpha$ is the fine-structure constant and  $n_e$ the electron density. While the photon coalescence is often neglected in massive axion production in the Sun, it turns out to significantly dominate over the Primakoff production for trapped axions due to the velocity-suppression effects \cite{DiLella2003, KKaxions}.

The production of trapped axions via the axion-electron coupling is dominated by the Compton mechanism for which we derive the following transition rate:
\be
\Gamma^{\rm{Comp.}} ~ \simeq ~ \alpha \gae^2 n_e ~\frac{m}{m_e^4}~ \sqrt{ m^2 - \omega_P^2} \,, \label{eq:ProdCompton}
\ee
where $\omega_P = \sqrt{4\pi\alpha n_e/m_e}$ is the plasma frequency. The Bremsstrahlung production described in \cite{VanTilburg2020} is also considered in the present work, although not detailed since it is always smaller than Compton production for $m \gtrsim 4$ keV.

The number density of trapped axions at a distance $\bar{r}$ from the Sun (normalized by the solar radius $R_\odot$) and a time $t$ is derived in \cite{KKaxions}:
\begin{widetext}
\begin{equation}
n(t, \bar{r}) ~=~ \bigg(2 \times 10^{13} {\rm cm}^{-3} \bigg)  \, \Big(1 - e^{-t\, \Gamma_{\agg}}\Big) \, \int_0^1 d \bar r_0 \bar r_0^2 ~ \frac{\Gamma_{\rm{Prod.}}}{\Gamma_{\agg}} \, \frac{m^3}{e^{m/T} - 1} ~ \frac{1}{4 \pi \bar r^4} \sqrt{2 \bigg(\bar \Phi(\bar r_0) - \frac{1}{\bar{r}}\bigg) }~,
\end{equation}
\end{widetext}
where
$ \bar \Phi$ is the gravitational solar potential normalized to its surface value, $m$ and $T$ are given in keV, and $\Gamma_{\rm{Prod.}}$ is a production transition rate that can be replaced by any of the ones described above.

The integral must be performed over a solar model. Since the uncertainties are not provided by any solar model, we compare three of them: the Saclay model \cite{SaclayModel}, the Vinyoles et al. model \cite{Vinyoles2016}, and the Bahcall et al. model \cite{Bahcall2004}. We assume that the theoretical uncertainties of the quantities derived in this work are given by the maximal differences obtained with the three solar models.

Finally, the axion-induced photons flux measured by a detector having a field of view noted $\alpha_0$ can be expressed as:
\be
\frac{dF_\gamma}{dE_\gamma} \simeq \Gamma_{\agg} R_\odot n(t_\odot, 1) \int_0^{\infty} d \bar{D} \int_{\cos \alpha_0}^1 \frac{d \cos \alpha}{\bar{r}^4}\,,
\label{eq:Flux}
\ee
where $\bar{D}$ is the normalized distance between the detector and the axion. This is the flux of photons of energy $E_\gamma = m/2$ measured by a detector such as SphinX or NuSTAR. The SphinX field of view is $\alpha_0 = 120~\rm{arcmin}$ whereas for NuSTAR we use a mean value of $\alpha_0 = 12~\rm{arcmin}$. An analytical expression of Eq.~\eqref{eq:Flux} is provided in \cite{KKaxions} for $\alpha_0 > \rm{atan} \big(R_\odot/\rm{AU}\big)$. 

\vspace{0.4cm}


\emph{The effect of the photon coalescence.---} In the following, we derive exclusion limits by requiring that the predicted axion-induced photons flux does not exceed the measured solar X-ray flux for each energy bin. This is a conservative method. Here we restrain the analysis to hadronic axions, \textit{i.e.} we set $\gae = 0$, in order to study the influence of the photon coalescence production on the exclusion limits.

The NuSTAR data used by DeRocco~\textit{et al.} are retrieved from their public Github \cite{Github} and then converted from a count per energy bin into a flux by dividing by the effective exposure time and by normalizing by the effective area, for each energy bin, according to the ancillary response functions that account for detector effects. No treatment is applied to the SphinX data since they are directly provided as a photon flux per energy bin. In the analysis, we assume that both telescopes are pointing towards the Sun since the solar shift during the measurements is comparable to the angular uncertainties. 

\begin{figure}[t]
\centering
\includegraphics[width=0.49\textwidth]{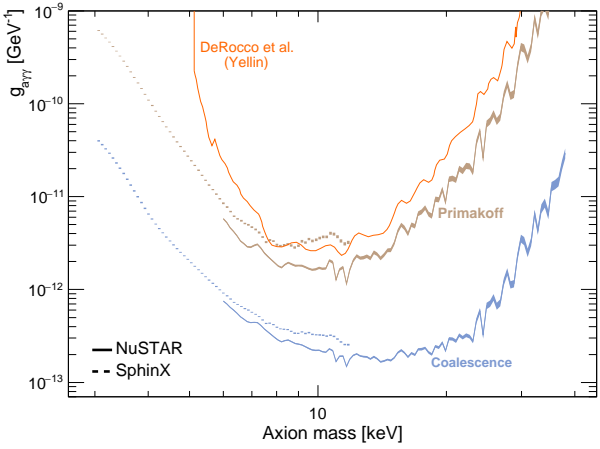}
\caption{Comparison of the exclusion limits in the parameter space $(\gagg, m)$ obtained when producing the axions by the Primakoff mechanism (brown lines) and by the photon coalescence mechanism (blue lines). The analysis is restrained to hadronic axions ($\gae=0$). The dashed curves are derived from SphinX data whereas the thick curves are derived from NuSTAR data. The thickness of the lines takes into account the uncertainties on the Solar model obtained by a comparison of three models. For comparison, the Yellin exclusion limit obtained by DeRocco~\textit{et al.} is represented in orange, based on the data available in \cite{Cajohare}.}
\label{fig:compareDeRocco}
\end{figure}

The influence of the coalescence production on the exclusion limits in the $(\gagg, m)$ parameter space, with respect to the Primakoff production, is represented in Fig.~\ref{fig:compareDeRocco}. Several comments are required at this stage. First, the addition of the photon coalescence in the axion production in the Sun improves by one order of magnitude the exclusion limits. Second, for the Primakoff mechanism, we obtain exclusion limits of the same order of magnitude than DeRocco~\textit{et al.}, demonstrating the consistency between our framework and the one developed in \cite{VanTilburg2020}. Third, the SphinX and the NuSTAR measurements are similarly restrictive, once accounting for the field of view, but they are complementary covering a larger parameter space. 

\vspace{0.4cm}


\emph{Compton absorption of trapped axions.---} The trapped axions cross the Sun during their orbits; the time spent in the solar interior as well as the depth reached being influenced by the initial conditions at production. A description of the equations of motion of the trapped axions can be found in section 3.4 of \cite{ThesisCyprien}. 

When crossing the Sun, the axion can be absorbed by a Compton-like scattering, $a + e^- \rightarrow \gamma + e^-$, whose cross-section in the non-relativistic limit is given by:
\be
\sigma_C ~=~ \frac{\alpha\gae^2 m^2}{m_e^4\,v}~,
\label{eq:ComptonCrossSec}
\ee
in which $v$ is the axion velocity and we have assumed that the axion energy is entirely transferred to the photon. We stress that this formula differs from the cross-section of axion emission by the presence of the velocity in the denominator. We will determine the Compton absorption probability by two complementary approaches: a Monte Carlo (MC) simulation and an analytical derivation. 

The probability for an axion to be absorbed by a Compton-like scattering over a time $T$ is given by: 
\be
    p_T = \int_0^Tdt\,\sigma n_e v \rightarrow \frac{\alpha\gae^2 m^2}{m_e^4}\lim_{n\to\infty}\sum_{i=0}^{n} n_e(t_i) \frac{T}{n}\equiv T\bar{\Gamma}~,
\ee
where we have used a Riemann sum and defined $\bar{\Gamma}$ to rewrite the expression more conveniently. On average, the axion has experienced a number $N = t_\odot/2T$ of periods T since its production. Assuming that $p_T$ is constant, so for large enough $T$, the mean absorption probability can be derived as:
\be
p^{\rm{tot.}} = 1 - \big(1 - p_T\big)^N \simeq 1 - \exp\big\lbrace-t_\odot\bar{\Gamma}/2\big\rbrace~.
\label{eq:MC}
\ee

The role of the MC consists in determining $\bar{\Gamma}$. We produce $n$ axions of mass $m$ in the Sun and some randomness is introduced in the choice of the initial conditions. Each axion is then tracked over $T\sim25~\rm{years}$ by solving the equations of motion, which allows determining $\bar{\Gamma}$ based on a solar model.

Let us now derive analytically the number density of trapped axions when considering Compton absorption, hereafter noted $n^C(t, \bar r)$. The time evolution of this quantity is governed by the Boltzmann equation:
\be
\frac{dn^C}{dt} ~=~ S ~-~ (\Gamma_C^{(v)}+ \Gamma_{\agg})n^C~,
\label{eq:Boltzmann}
\ee
where $S$ is a source term of trapped axions and $\Gamma_C^{(v)}$ the Compton absorption rate whose superscript indicates that it depends on the velocity distribution of the trapped axions. 

We start by expressing the source term:
\be
S ~=~ \frac{1}{(2 \pi)^3} 4 \pi R_\odot^3 \int d \bar r_0 \bar r_0^2 \int d^3 p P_{\bar{r}}(v) \frac{\Gamma_{\rm{Prod.}}}{e^{m/T} - 1} ~,
\ee
where one integral runs over a solar model and the other over momenta, and where we introduced $P_{\bar{r}}(v)$ the probability density of a trapped axion to be at a distance $\bar r$ from the Sun such that:
\be
\int_{Sun} d^3 \bar r P_{\bar r}(v)= 1 \,.
\ee

We now introduce another probability density, $P_T(v)$, describing its velocity distribution and for which the normalization reads:
\be
\int d^3 v P_T(v) =1 \,.
\ee
We know from Eq.~\eqref{eq:ProdCompton} that the Compton transition rate does not depend on the velocity but that it indirectly depends on the radius inside the Sun through the electron number density. The Compton absorption rate is then given by:
\be
\Gamma_C^{(v)} = 4 \pi R_\odot^3 \int d \bar r_0 \bar r_0^2  \Gamma^{\rm{Comp.}}(\bar r_0) \int d^3 v P_T(v) P_{\bar r}(v)
\ee

\begin{figure}[t]
\centering
\includegraphics[width=0.49\textwidth]{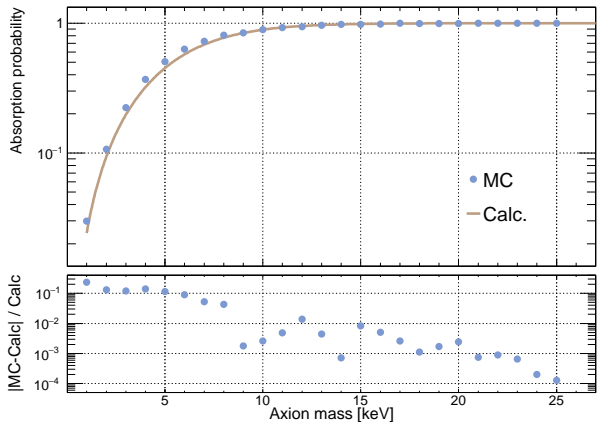}
\caption{Top: the absorption probability obtained from MC and the analytical calculation for $\gae = 10^{-12}$ and $\gagg = 6\times 10^{-13} ~\rm{GeV^{-1}}$. Bottom: the residual quantifying the agreement between the two approaches.}
\label{fig:comMCcalc}
\end{figure}

From our previous work \cite{KKaxions}, we know that:
\be
P_{\bar r}(v) = \left(\frac{GM_\odot}{R_\odot}\right) \frac{1}{32 \pi R_\odot^3} \Big(2 \bar \Phi_G(\bar r_0) -\bar{v}^2\Big)^4 \delta\big(f_T(v)\big)~,
\ee
with
\be
f_T(v, v_\phi) = v^2 - v_\phi^2 \Big(\frac{\bar r_0}{\bar r}\Big)^2 -\frac{2 GM_\odot}{R_\odot} \Big( \bar \Phi_G(\bar r_0) - \frac{1}{\bar r}\Big)~,
\ee
where we defined $\bar v = vR_\odot/(GM_\odot)$. For radial trajectories, 
the Compton absorption rate is given by:
\be
\Gamma_C^{(v)} = \frac{\int d \bar r_0 \bar r_0^2 ~\Gamma^{\rm{Comp.}} ~ \bar v~ P_T( \bar v )}{\int d \bar r_0 \bar r_0^2 ~  \bar v ~ P_T( \bar v)}~,
\ee
with ${\bar v}^2=2 (\bar \Phi_G(\bar r_0) - 1/\bar r)$. Finally, the last step consists in determining the probability density of the velocity distribution $P_T(v)$. The MC simulation tells us that it can be approximated by a Landau distribution. Note that for simplicity one could make a Gaussian ansatz, resulting in similar exclusion limits. The Landau distribution $\Phi_L(\bar v, \mu, \sigma)$ is given by:
\begin{align}
\Phi_L(\bar v, \mu, \sigma) &= \frac{p(\lambda)}{\sigma} \,,\;\;\; \lambda= (\bar v - \mu)/\sigma \,, \\
p(\lambda)&= \frac{1}{\pi} \int_0^\infty dt e^{(-t \ln t - \lambda t)} \sin(\pi t) \,,
\end{align}
with $\mu \simeq 0.08$. 
The coefficient $\sigma$ shows some mild dependence on the axion mass, varying between $\sigma \simeq 0.2$ for low masses and $\sigma \simeq 0.1$ for $m \gtrsim 20$ keV. We have then used  the interpolating function:
\be
\sigma \simeq a_0 \left ( 1 + \frac{1}{1+e^{(m-a_1)/a_2}} \right) \,, 
\ee
with $a_0 =0.1$, $a_1=15$, $a_2=4$ and $m$ given in keV. A comparison between the absorption probability (neglecting the decay) obtained by MC, Eq.~\eqref{eq:MC}, and the one derived analytically, $1 - e^{-\Gamma_C^{(v)}\,t}$, is presented in Fig.~\ref{fig:comMCcalc}. The agreement between the two approaches improves as the mass increases, as shown by the plot of the residual, with a difference smaller than $5\%$ for $m\geq 7~\rm{keV}$. Such a good agreement acts as a robust cross-check for the analytical derivation of the absorption probability.  

We now have all elements to express $n^C(t, \bar r)$ by integrating the Boltzmann equation:
\be
n^C(t, \bar r) = St~ \frac{1 - e^{-(\Gamma_C^{(v)}+\Gamma_{\agg})\,t}}{(\Gamma_C^{(v)}+\Gamma_{\agg})\,t} ~,
\ee
which can be introduced into Eq.~\eqref{eq:Flux} to obtain the axion-induced photon flux when considering Compton absorption. Similarly than for the production of trapped axions, the Compton absorption and axion decay embedded into the quantity $n^C(t, \bar r)$ significantly dominate over the other absorption mechanisms.

\vspace{0.4cm}


\emph{New exclusion limits from solar X-ray measurements.---} Hadronic axions are not affected by Compton absorption so the new exclusion limits in the parameter space $(\gagg, m)$ are the ones presented in Fig.~\ref{fig:compareDeRocco}. Note that the limits derived in this work do not rely on the usual assumption that ALPs form the entire dark matter of the galaxy. In this sense, we show the currently leading limits between $3~\rm{keV}$ and $40~\rm{keV}$ not relying on any assumption about the local dark matter density.

\begin{figure}[t]
\centering
\includegraphics[width=0.49\textwidth]{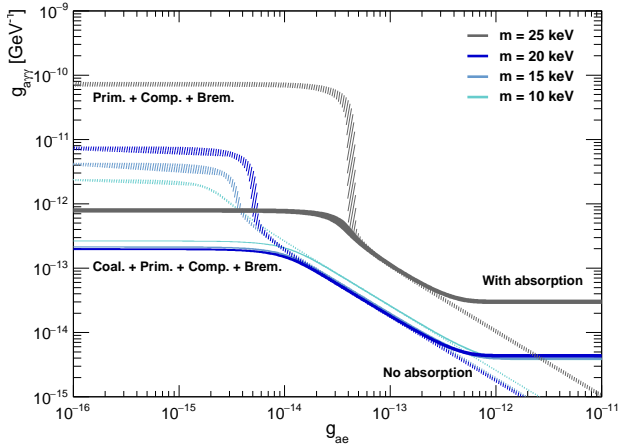}
\caption{Exclusion limits in the parameter space $(\gagg,\, \gae)$ for multiple axion masses as indicated in the plot, from combined solar X-ray measurements of SphinX and NuSTAR. The thick lines correspond to all production mechanisms and account for Compton absorption which counter-balances the production for $\gae \gtrsim 10^{-12}$. The dashed lines ignore the axion production via the photon coalescence as well as Compton absorption. The thickness of the lines takes into account the theoretical uncertainties.}
\label{fig:gayy_gae}
\end{figure}

The relative influences of the production mechanism and the absorption are represented in Fig.~\ref{fig:gayy_gae}. When the coalescence production is neglected, a clear separation can be observed between a region governed by the Primakoff production for small $\gae$, and a region dominated by Compton production. The abrupt change of regime is enhanced by the decay of the axions related to the Boltzmann equation. The situation differs however when including the more efficient coalescence production for trapped axions, which sets lower limits on $\gagg$ for negligible values of the axion-electron coupling as discussed previously.

\begin{figure}[t]
\centering
\includegraphics[width=0.49\textwidth]{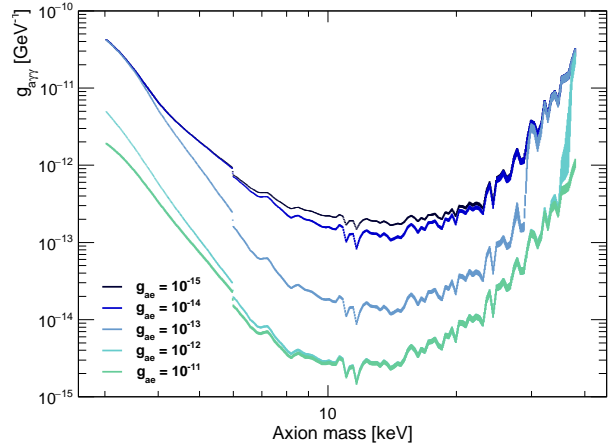}
\caption{Exclusion limits in the parameter space $(\gagg, \,m)$ for multiple values of $\gae$ as indicated in the plot, from combined solar X-ray measurements of SphinX and NuSTAR. We have included all processes considered in this work: Coalescence, Primakoff, Compton, and Bremsstrahlung, as well as Compton absorption. The thickness of the lines takes into account the theoretical uncertainties.
}
\label{fig:gayy_ma_gae}
\end{figure}

The main comment concerns the crucial role played by Compton absorption as a counter-balance of the non-hadronic axion production. One can identify two regimes for which the limits are exclusively governed by $\gagg$: one for low $\gae$ where the Compton production and absorption are negligible; the other for large $\gae$ for which the absorption entirely compensates the production. In Fig.~\ref{fig:gayy_ma_gae}, we have plotted the same limits including all the processes and Compton absorption, but now in the plane $(\gagg, \,m)$ for different values of $\gae$. For masses $m \lesssim 20$ keV, the transition from the upper to the lower bound on $\gagg$ starts around $\gae \simeq 10^{-14}$, while for larger masses this is shifted toward larger values of $\gae$. 

The exclusion limits published in \cite{VanTilburg2020} in the parameter space $(\gae, m)$ from direct detector measurements are negligibly affected by Compton-absorption because the trapped axions of interest in the analysis are located near the Earth. With such large orbits, the typical gravitational ejection time drops below the age of the Sun, so it can no longer be neglected, and the frequency of Sun crossings significantly decreases. Finally, we mention that a recent work on the interpretation of the axion-electron coupling \cite{Smith2023} could affect the phenomenology of non-hadronic axions.

\vspace{0.4cm}


\emph{Conclusion and discussions.---} 
The Sun represents a particularly interesting source for keV ALPs since a fraction of them would be gravitationally trapped and accumulate over cosmic times. The comparison of solar X-ray measurements with the predicted photon flux of decaying trapped ALPs set constraints on ALP models, significantly improved compared to the non-trapped case, without relying on any assumption about the local DM density, like those from X-rays observations from the Milky Way or the galactic halo \cite{Ng:2019gch, Roach:2019ctw,Foster:2021ngm, Roach:2022lgo}. Even if the ALP does not account for the total DM abundance, it has been pointed out in \cite{Langhoff:2022bij} that there is always an "irreducible" background contribution coming from the freeze-in mechanism at early times. This translates into stringent limits on their couplings from the non-observation of their decay in X-ray data, assuming that the irreducible ALP background follows the same local distribution than the DM one. Given the uncertainties in the later, we have derived a similar constraint but directly from solar measurements.

In this letter, we have stressed the importance of considering the ALP production via the photon coalescence which improves by one order of magnitude the existing limits. When its orbit crosses the Sun, an ALP could be absorbed by a Compton-like scattering. When integrated over the ALP lifetime, such an absorption significantly counter-balances Compton production. In the exclusion limits, Compton absorption results in two well-defined regimes whose amplitudes are exclusively governed by $\gagg$, with a transition determined by $\gae$.

The indirect detection of trapped ALPs, or stringent exclusion limits, could be achieved from the typical $1/r^4$ tendency of the number density where $r$ is the distance to the Sun. Measurements of the photons flux at multiple distances to the Sun, such as the ones currently performed by the STIX telescope \cite{STIX}, could exhibit such a behavior. 

Finally, let us mention that the X-ray solar spectra measured by SphinX or NuSTAR significantly deviate from theoretical predictions based on element abundances above $2.5~\rm{keV}$ \cite{Sylwester2019}. While the decays of trapped ALPs into photons would result in a line in an X-ray spectrum, the decays of higher-dimensional axions would form a continuum whose predicted spectrum relatively matches the measurements \cite{KKaxions}, possibly explaining the deviation to the isothermal distribution observed in the solar X-ray spectrum. 

\vspace{0.4cm}

\begin{acknowledgments}
\emph{Acknowledgments.---} We thank Barbara and Janusz Sylwester who kindly shared and discussed the SphinX measurements with us.
\end{acknowledgments}

\bibliographystyle{apsrev4-1}
\bibliography{references}

\end{document}